\documentstyle[12pt,bezier]{article}
\begin{document}
\setlength{\baselineskip}{0.27in}

\newcommand{\beq}{\begin{equation}}
\newcommand{\eeq}{\end{equation}}
\newcommand{\beqa}{\begin{eqnarray}}
\newcommand{\eeqa}{\end{eqnarray}}
\newcommand{\lsim}{\begin{array}{c}\,\sim\vspace{-21pt}\\< \end{array}}
\newcommand{\gsim}{\begin{array}{c}\sim\vspace{-21pt}\\> \end{array}}
\newcommand{\nt}{\nu_\tau}
\newcommand{\nee}{\nu_e}
\newcommand{\nm}{\nu_\mu}
\newcommand{\bi}{\bibitem}

\begin{titlepage}
\pagestyle{empty}
\rightline{June, 1994}
\rightline{UM-TH-94-21}
\rightline{UMN-TH-1303-94}
\rightline{FERMILAB-Pub-94/199-A}
\rightline{hep-ph/yymmddd}
\begin{center}
\vglue .06in
{\Large \bf Bounds on Dirac Neutrino Masses \\
from Nucleosynthesis} \\[.5in]

A.D. Dolgov$^{1,2,3}$,
K. Kainulainen$^4$
and I.Z. Rothstein$^3$
\vskip 1truecm

$^1${\it Institute of Theoretical and Experimental Physics,}\\
{\it Moscow 117259, Russia,}\\
$^2${\it NASA/Fermilab Astrophysics Center, Fermi National} \\
{\it Accelerator Laboratory, Batavia, IL 60510 }\\
$^3${\it Randall Laboratory of Physics,} \\
{\it University of Michigan, Ann Arbor, MI 48109-1120}\\
$^4${\it School of Physics and Astronomy, University of Minnesota}\\
{\it Minneapolis, MN 55455, USA}
\vskip 1truecm

{\bf Abstract}\\[-.05in]

\begin{quote}
We derive new bounds on the Dirac mass of the tau and muonic
neutrinos. By solving the kinetic equation for the rate of
energy deposition due to helicity flipping processes and
imposing the constraint that the number of effective species
contributing to the energy density at the time of nucleosynthesis
be $\Delta k_\nu<~0.3$, we find the
bounds $m_{\nu_\mu} < ~150$ KeV and  $m_{\nu_\tau} < ~190$ KeV
for $T_{\rm QCD}= 200$ MeV.
The constraint $\Delta k_\nu~<0.1~$  leads to the much
stronger bound $m_\nu <10$ KeV for both species of
neutrinos.
\end{quote}
\end{center}

\end{titlepage}
\newpage

If neutrinos are massive and stable, then the mass range
$40~{\rm eV} < m_\nu < 2$ GeV \cite{gz,hut} is excluded by measurements
of the age of the universe. However, if the neutrinos are unstable then
these bounds will not apply and we must search elsewhere to find
mass restrictions. The present laboratory bounds on the muonic and tau
neutrino are $160$ KeV\cite{r} and $31$ MeV \cite{argus} respectively.
Better bounds may be had if the lifetime of the massive neutrino
is $>{\cal O}(100)$ sec. In this range of lifetimes the energy density
due to the massive species may come to increase the expansion rate
of the universe and thereby increase the primordial helium
abundance \cite{ks}. In this case the mass ranges
$0.5 < m <~ 35$ MeV  and $0.3 < m <~ 35$ MeV were found to
be excluded for Majorana and Dirac neutrinos respectively
\cite{dr}. The mass bound for Dirac neutrinos was
derived assuming that the right handed species is in thermal
equilibrium below the QCD phase transition, for
$m > 300$ KeV. In this letter we show that even if the
right handed neutrinos do not enter thermal equilibrium at
temperatures below $T_{QCD}$, their out-of-equilibrium production
rate is strong enough to yield a much more stringent lower bound
on the Dirac neutrino mass.

The rate of  production of right handed neutrinos is  roughly
given by, $\Gamma_- \propto (m_\nu /E)^2 G_F^2T^5$ where $E$ is
the energy of the neutrino and $\Gamma_W$ is the normal weak interaction
rate.  Thus, the ``wrong'' helicity states\footnote{In the rest of
the paper, when we refer to ``wrong'' helicity states, we mean right
handed neutrinos and left handed anti-neutrinos.} are produced more
efficiently at higher temperatures. If the wrong helicity states
decouple from thermal equilibrium above $T_{\rm QCD}$, their energy
density will be greatly diluted in comparison with the coupled
species as a consequence of entropy conservation. Such neutrinos
would appear to be cosmologically safe. A bound based on the
decoupling argument was first derived in
ref.\ \cite{fm}, where it was shown that the Dirac mass must
satisfy $m_\nu < 300$ KeV, assuming that the  QCD phase transition
temperature is 100 MeV. The result of ref.\ \cite{fm}, was
refined in ref.\ \cite{eu}, where a more careful computation of
the relevant scattering processes was performed and much weaker
bounds were claimed.

In ref.\ \cite{ln} a different bound, $m_{\nu_{\mu}}<420$ KeV was
obtained by considering the out-of-equilibrium production of
wrong helicity states through the pion resonance
$\gamma \gamma \rightarrow \pi^0 \rightarrow \nu_+ {\bar \nu}_+$,
and imposing the constraint $\Delta k_\nu<0.3$. Unlike the bound
discussed first above, this result is insensitive to $T_{\rm QCD}$,
as will be shown below.

In this letter we give a more accurate analysis of the production
of light wrong helicity states in the early universe. We will account
for all sources of wrong helicity neutrinos, including the population
that decoupled above $T_{\rm QCD}$ as well as those generated
from the following out-of-equilibrium processes:
\beqa
\pi^0 &\rightarrow& \nu_{\mu(\tau)+}
{\bar \nu}_{\mu(\tau)+}  \nonumber \\
\pi^+ &\rightarrow& \mu^+ \nu_{\mu+}  \nonumber \\
\label{intact}
l_1l_2& \rightarrow & l_3 \nu_{\mu(\tau)+}\\
l{\bar l}& \rightarrow &\nu_{\mu(\tau)+}{\bar \nu_{\mu(\tau)+}}.
\nonumber
\eeqa
Moreover, we will accurately solve the kinetic equation for neutrino
energy density generated out-of-equilibrium at temperatures below
the QCD phase transition. Our bounds will be both more precise and
much stronger than any found in earlier literature.

We should mention here that strong bounds on the neutrino mass,
$m_\nu <10-20$ KeV, were obtained from the consideration of the cooling
of supernova SN-87 \cite{mzy,bgt}. While these bounds are indeed more
stringent than the bounds obtained in this paper, they are in a sense
model dependent. Supernova bounds come from calculating the rate
at which energy is drained from the supernova assuming that the right
handed neutrinos are sterile and will free stream from the core.
However,
since the right handed neutrinos are singlets, it is quite
possible that they interact in yet unknown ways
\cite{bmr1}.\footnote{ If the right handed species is trapped
inside supernova, thus obviating the bound derived in ref.\
\cite{mzy,bgt}, it would likely contribute to the energy
density at the time of nucleosynthesis. To avoid this, physics
far beyond the standard model is needed \cite{bmr2}.}
Furthermore, with better data it may be possible that the
nucleosynthesis bounds will eventually become more stringent
than those obtained from supernova considerations.
\vskip0.3truecm

We will first consider helicity flipping scatterings
$\nu_{\mu(\tau)-} l \rightarrow \nu_{\mu(\tau)+} l$, and annihilations
$l {\bar l} \rightarrow  \nu_{\mu(\tau)+}{\bar \nu_{\mu(\tau)+}}$
where $l$ is a lepton (not the tau however)
and sub-$\pm$ refers to the helicity of the neutrino.
The amplitude \ for $t$-channel elastic
scattering process is given by
\beq{
A(\nu l; \nu_+l) = {G_F\over \sqrt 2} \bar \nu \gamma_\alpha (1-
\gamma_5) \nu_+ \bar l (c_V - c_A\gamma_5)\gamma_\alpha l,
}\label{ampl}
\eeq
where, for example in $\nu\nu$-scattering $c_V = c_A={1\over 2}$ and
in $\nu e$-scattering $c_A=2\sin^2\theta_W-{1\over 2}$ and
$c_V= {1\over 2}$. The projection onto the wrong helicity
state is accomplished by using
\beq{\sum_{spins}
\nu_+ {\bar \nu_-} = {1\over{2}}(k'\!\!\!\! \slash + m_\nu  )
(1+\gamma_5 \hat{s}),
}\label{spin}
\eeq
where the four-vector $s$ is  given by the expression:
\beq{
s= {\lambda \over m_\nu} (k', \omega' \vec{n'})
.}\label{s}
\eeq
Here the helicity eigenvalue $\lambda =\pm 1$ determines the helicity state
of final state neutrino, $k'=|\vec{k'}|$ is the magnitude of the spatial
component of the neutrino momentum, $\omega ' = (k'^2 +m_\nu^2)^{1/2}$ is
its energy, and $\vec{n}$ is the unit vector in direction of $\vec{k'}$.

The amplitude squared for the production of the neutrinos with the
wrong helicity in elastic scattering is
\beqa
|A(\nu l; \nu_+ l)|^2 &=&16 G_F^2 \{(c_V\pm c_A)^2(p\cdot k)(p'\cdot l)
  \nonumber \\  &&\phantom{han}  + (c_V\mp c_A)^2(p'\cdot k)(p\cdot l)
                                 - (c_V^2 - c_A^2) m_l^2(k\cdot l)\}
\label{awrong}
\eeqa
where $p$, $p'$,  are the four-momenta of
the initial and final fermions, and $k$, $k'$  are the momenta of
the neutrinos $\nu_-$
and $\nu_+$ respectively. The upper (lower) signs refer to
scattering off particles (antiparticles) and we defined the
four-vector $l= (\omega ' -k') (1, -\vec{n}) =
(m^2_\nu /2k ') (1, -\vec{n})$. The amplitude for annihilation
process may be obtained from  (\ref{awrong}) by simple crossing
relation. One may see that for the scattering off of
neutrinos, $|A|^2$ goes like the neutrino mass to the fourth power
in the center of mass frame. It is because of this seemingly slow
rate that these processes were neglected in ref. \cite{eu}. However,
this suppression is not present in an arbitrary frame and generically
the production of the wrong helicity neutrinos is suppressed only
as $m_\nu^2$. This is a consequence of the fact that the helicity is
not a Lorentz invariant quantity. Indeed,
given a particle with definite negative helicity, a boost
in a direction orthogonal to its direction of motion with velocity $v$
will generate an admixture of the positive helicity equal to $m^2v^2/4E^2$.

The kinetic equation for the energy density of the right-handed
neutrinos has the form
\beqa
{d\rho_\nu \over dt} + 4H\rho_\nu &\equiv& C_t~=~ {1\over (2\pi)^8}
\int {d^3p\over 2E}{d^3k \over 2\omega}
{d^3p'\over 2E'}{d^3k' \over 2 \omega'}\omega'
\delta^4(p+k-p'-k') \times \nonumber \\
& &\phantom{hannatytto} \times \left\{ \; |A_t|^2 f_\nu (\omega)
f_l(E) [1-f_{l}(E')] + ... \right\},
\label{kineq}
\eeqa
where the ellipses refer to the remaining elastic scattering
(off antiparticles) and annihilation channels as well as to the
contributions from the pion resonances which will be treated below.
Note that we omitted the Pauli blocking factor for wrong helicity
states, which are {\it not} in thermal equilibrium.
We also neglected the inverse reaction of right-handed neutrinos
producing the normal ones. This is justified because the density
of the wrong neutrinos has to be small. For the same reason we
can compute all contributions to $\rho_\nu$ separately and
add the different contributions in the end.

If we assume that the phase space distribution functions are of
the equilibrium Boltzmann form, $f=\exp(-E/T)$, we can compute
the collision integral on the r.h.s.\ of equation (\ref{kineq})
analytically in the massless limit. For example, the matrix element
for a $t-$channel reaction in (\ref{awrong}) gives
\beq
C_t(m_f=0)_{MB} = \frac{G_F^2 m_\nu^2 T^7}{16\pi^5}
      \{ (c_V+c_A)^2 + \frac{5}{6}(c_V-c_A)^2 \}
\label{onechannel}
\eeq
We have also computed the collision integrals numerically retaining
the effect Fermi statistics, and found a 21\% suppression
relative to the case of Boltzmann statistics. Furthermore, it
was found that this suppression is nearly the same for all the
reactions.

We took into account the following scattering processes:
$\nu_{\mu-} \nu_e \rightarrow \nu_{\mu+} \nu_e ,\;
\nu_{\mu-} \nu_\mu \rightarrow \nu_{\mu+} \nu_\mu ,\;
\nu_{\mu-} \nu_\tau \rightarrow \nu_{\mu+} \nu_\tau ,\;
\nu_{\mu-} e^- \rightarrow \nu_{\mu+} e^- ,\;
\nu_{\mu-} \mu^- \rightarrow \nu_{\mu+} \mu^- ,\;$
with the corresponding channels for scattering off antiparticles
and the annihilation channels. Similar channels exist also for
$\nu_{\tau+}$-production. However, the reaction channels involving
muons are different, due to the contribution from charged current
scattering. Moreover, in the muonic
case there are three more pure charged current channels that do
not exist for $\nu_{\tau+}$-production: $\mu^- \nu_e
\rightarrow \nu_{\mu+} e^- ,\;
\mu^- e^+ \rightarrow \nu_{\mu+} \bar{\nu_e} $ and
$e^+\nu_e  \rightarrow \nu_{\mu+} \mu^+$. There is also a contribution
from scattering off of pions, which we found to be small and will
be neglected.

Equation (\ref{onechannel}), modified to include Fermi statistics,
 can be used
for all relevant reaction channels except for those involving muons.
For  the muonic channels we computed the collision integrals numerically
as a function of temperature. Altogether we can write the kinetic
equation as
\beq
\dot \rho_\nu +4H\rho_\nu =
\frac{G_F^2m_\nu^2}{2\pi^5}c_{\rm F}N_{\rm eff}T^7,
\label{dot rho}
\eeq
where the Fermi correction factor $c_{\rm F} \simeq 0.79$ and the
effective number of reactions $N_{\rm eff}$ (normalized such that
$t$-channel scattering off neutrinos gives a contribution 1),
for example in $\nu_\mu$-case is given by
\beq
N_{\rm eff,\mu} = (\frac{16}{3}) \cdot
\{1 + \sum_{f = e,\nu_e, \nu_\tau}
(c_{Vf}^2+c_{Af}^2)  + (c_{V\mu}^2 +c_{A\mu}^2 + 2)d_\mu(T) \}.
\eeq
The first '1' comes from scattering off of muonic neutrinos and
\ $c_{V\mu}=2\sin^2\theta_W+\frac12$, $c_{A\mu}=\frac12$
(see also below equation (\ref{ampl})). Using $\sin^2\theta_W = 0.23$,
we find
\beqa
N_{\rm eff,\mu} &\simeq& 3.92 (1 + 0.81d_\mu(T)) \nonumber \\
N_{\rm eff,\tau} &\simeq& 3.92 (1 + 0.06d_\tau(T)),
\label{fits}
\eeqa
where the fit functions $d_i(T)$ are normalized to become unity for
very large $T$. The suppression due to the muon mass becomes significant
rather late: for $T=m_\mu/3$ we still get $d_\mu \simeq 0.42$ and
$d_\tau \simeq 0.35$. Hence the muonic interactions are important
in deriving the mass bounds. For $T=m_\mu/10$ however, we find
$d_\mu \simeq 4.7\times 10^{-3}$, which clearly validates our
neglect of reactions involving $\tau$s.

If all the particles in the plasma were massless we could easily
integrate the equations (\ref{dot rho}-\ref{fits}) since in this case
the temperature and the time are related as $T^2t=const$. However,
since the contribution of massive pions and muons to the energy
density is essential, we should proceed more carefully.
If we neglect the tiny contribution to the pressure density $p$
coming from the wrong helicity states, the covariant energy
conservation law $\dot \rho =-3H(\rho +p)$ reduces to the usual
conservation of entropy of interacting species, and we get
the standard relation between time and temperature
\cite{osw}
\beq{
{dt \over dT} = - ({45\over 4\pi^3})^{1/2}
{m_{Pl}\over T^3} {1 \over g_*^{1/2}(T)}
\left(1+{T\over 3h_*}{dh_*\over dT}
\right).
}\label{dt}
\eeq
Here $g_*$ and $h_*$ are the effective numbers of energy and
entropy degrees of freedom, defined through the total energy density
$\rho \equiv {\pi^2 \over 30} g_*(T) T^4$ and the entropy density
of interacting species $s_I \equiv {2\pi^2 \over 45} h_*(T) T^3$.

Equations (\ref{dot rho}-\ref{dt}) are easily solved to yield
a simple integral expression for the relative energy density
$r_1\equiv \rho_{\nu+}/\rho_{\nu-} =
\rho_{\nu+}/(7\pi^2 T^4 /240)$:
\beq{
r_1 = 2.88
\left({ m_\nu \over{\rm MeV}} \right)^2
\int^{T_{\rm QCD}}_0 {\rm d}T
\left( {10.75 \over h_*(T) }\right)^{4/3}
\left( {10.75 \over g_*(T) }\right)^{1/2}
\left( 1+ {T\over 3h_*} { dh_* \over dT}\right) (1 + a_id_i(T)).
}\label{r1}
\eeq
Sub-1 means that we take into account only weak scatterings; pion
decays are treated separately below. $T_{\rm QCD}$ is the temperature
of the QCD phase transition and the coefficients $a_i$ can be read off
the equation (\ref{fits}), $a_\mu=0.81$ and $a_\tau=0.06$.
Equation (\ref{r1}) incorporates the
effect of the changing energy density due to annihilations of heavy
particles, as well as the statement of entropy conservation.
We have integrated (\ref{r1}) numerically and found that (for $T_{\rm QCD}
\gsim 100$) to a very good accuracy $r_i$s follow the simple linear
fits
\beqa
r_{1\mu} &=&
(0.53 + 2.25 \; T_{\rm QCD}^{100})\; (m_\nu /{\rm MeV})\nonumber \\
r_{1\tau}&=&
(0.08 + 1.34 \; T_{\rm QCD}^{100})\; (m_\nu /{\rm MeV}),
\label{rfits}
\eeqa
where $T_{\rm QCD}^{100}$ is the QCD phase
transition temperature in units 100 MeV.
A word of caution is in order here: In equation (\ref{r1}) we assumed
that $g_*$ and  $h_*$ reach a constant value 17.25 well above pion
mass.  This indeed seems to be the correct choice \cite{osw}, but
should they instead increase significantly at temperatures between
$m_\pi$ and $T_{\rm QCD}$, then equations (\ref{rfits}) would slightly
overestimate $r_1$ for $T_{\rm QCD} \gsim {\cal O} (200)$ MeV.

The decays $\pi^0 \rightarrow \nu \bar \nu$ and $\pi^{\pm} \rightarrow
\mu\nu_\mu$ also  give significant contributions to the energy density
of right-handed neutrinos. The decay of $\pi^0$ produces both
$\nu_\mu$ and $\nu_\tau$ while $\pi^{\pm}$ produces only $\nu_\mu$.
The former was considered in ref.\ \cite{ln} while the latter, to
best of our knowledge, has not been accounted for to date. Both of these
processes have the property that their contribution  to the
energy density is insensitive to the temperature of the QCD transition
because the dominant contribution to production of the right-handed
$\nu$ occurs at $T\approx m_\pi /5$. Let us first consider the decay
$\pi ^0 \rightarrow \nu_+ {\bar \nu}_+$. This width was calculated in
ref.\ \cite{fish,ksv} as
\beq{
\Gamma (\pi^0 \rightarrow  \nu_+ {\bar \nu}_+) =
G^2_F f^2_{\pi^0} m_\nu^2 m_\pi /8\pi,
}\label{gamnu}
\eeq
where $f_{\pi^0} \simeq 93$ MeV. When computing the $\pi^0$ collision
term, we will assume the Maxwell-Boltzmann statistics. This is an
excellent approximation here and moreover we are able to compute
the collision term analytically.  We find that the energy density
of neutrinos produced in $\pi^0$-decays satisfies the equation
\beq{
\dot \rho_\nu + 4H\rho_\nu = \frac{G_F^2m_\nu^2}{32\pi^3}
f_{\pi^0}^2T^5x_0^4K_2(x_0),
}\label{rho0}
\eeq
where $x_0 \equiv m_{\pi^0}/T$ and $K_2(x)$ is the usual modified Bessel
function. This equation can be integrated analogously to the scattering
contribution (cf.\ equation (\ref{r1})) with the result
\beq{
r_2 = 1.21 (m_\nu/{\rm MeV})^2,
}\label{r2}
\eeq
which to the given accuracy is independent of $T_{\rm QCD}$.
Quite similarly we find for the decay $\pi^+\rightarrow \mu^+ \nu_+$:
\beq{
\dot \rho_\nu + 4H\rho_\nu = \frac{G_F^2m_\nu^2}{32\pi^3}
f_{\pi^\pm}^2T^5x_+^4 (1-y^2) \{ (1+y^2) K_2(x_+) - y^2 K_0(x_+)\},
}\label{rhopm}
\eeq
where $f_{\pi^\pm} \simeq 128.4$, $x_+ \equiv m_{\pi^\pm}/T$ and
$y\equiv m_\mu/m_{\pi^\pm}$. When integrated, (\ref{rhopm}) gives
\beq{
r_3 = 1.15 (m_\nu / {\rm MeV})^2.
}\label{r3}
\eeq

At temperatures above the QCD phase transition,
the  plasma consisted of free quarks and gluons
in addition to leptons and photons and the effective number of degrees
of freedom $g_*$ was slightly below 60. Since the efficiency of
production of right-handed neutrinos is larger at higher temperatures
(at least up to the temperatures of the order of the $W$ and $Z$
boson masses) we expect them to be in thermal equilibrium if their
mass is above ${\cal O} (10)$ KeV. The energy density of these neutrinos
will be diluted in the course of QCD-phase transition  and
due to annihilation of massive states in the plasma.
Using the entropy conservation we may estimate their
relative energy density as
\beq{
r_{\rm QCD}= (10.75 /60)^{4/3} \simeq 0.10.
}\label{rqcd}
\eeq

There is one additional effect that needs to be taken into account.
Namely, the fact that energy density of a massive species will scale
differently than that of a massless species. Thus, the there will
be additional contributions to the relative energy density
coming from population of left handed neutrinos as well as the small
population of right handed neutrinos. Therefore, the total relative
energy density may be written as
\beq
r_{\rm {tot,\nu}}=f(m_\nu)+(\sum_i r_{i,\nu}+0.1)\; (1+f(m_\nu)).
\label{total}
\eeq
The function $f(m_\nu)$ was found using the results from paper
\cite{dr}, and is given by
\beq
f(m_\nu) = 1.31m_\nu^2 + 4.84 m_\nu^4  - 4.57 m_\nu^6
\equiv 1.31m_\nu^2 + \delta f,
\label{rm}
\eeq
and $m_\nu$ is given in units of MeV.
In this way we find
the bounds
\beqa
m_{\nu_\mu} &<& \left( {\Delta k_\nu^{\rm max} - 0.10
- \delta_\mu  \over
4.33 + 2.25 T_{\rm QCD}^{100}} \right)^{1/2} {\rm MeV},
\label{fbmu}
\eeqa
\beqa
m_{\nu_\tau} &<& \left( {\Delta k_\nu^{\rm max} - 0.10
- \delta_\tau \over
2.73 + 1.34 T_{\rm QCD}^{100}} \right)^{1/2} {\rm MeV},
\label{fbtau}
\eeqa
where $\delta_\mu = 1.10\delta f +(2.89 + 2.25 T_{\rm QCD}^{100})f$
and $\delta_\tau = 1.10\delta f +(1.29 + 1.34 T_{\rm QCD}^{100})f$.
Due to smallness of $\delta_i \sim {\cal O}(m_\nu^4)$, the
constraint equations (\ref{fbmu}) and (\ref{fbtau}) are easily solved
for any values of $\Delta k_\nu$ and $T_{\rm QCD}$. For example,
imposing the constraint $\Delta k_\nu < 0.3$
\cite{nuc1}, leads to the bounds
\beq
\label{eq:bounds}
m_{\nu_\mu} \: \lsim \:
\left\{
\begin{array}{lll}
&170~{\rm KeV} &
\;  \; T_{QCD}=100~{\rm MeV} \\
&150~{\rm KeV} &
\;  \; T_{QCD}= 200~{\rm MeV} \nonumber
\end{array}
\right. \,
\eeq
\beq
\label{eq:bounds2}
m_{\nu_\tau} \: \lsim \:
\left\{
\begin{array}{lll}
&210~{\rm KeV} &
\;  \; T_{QCD}=100~{\rm MeV} \\
&190~{\rm KeV} &
\;  \; T_{QCD}= 200~{\rm MeV.} \nonumber
\end{array}
\right.
\eeq

Let us finally mention that there were claims recently that the effective
number of extra neutrino species is not larger than 0.1 \cite {so}.
If this is indeed the case the right-handed neutrinos should
decouple before or near electroweak phase transition. Using this bound
the limit on both the $\nu_\tau$ and $\nu_\mu$  masses is,
conservatively, $10$ KeV.

\vskip 0.3truecm
A.D.Dolgov is grateful for hospitality to the Department of
Physics of University of Michigan where this work started
and to the Astrophysical Group at Fermilab where it was finished.
K.K.\ wishes to thank the Finnish Academy for financial support.
This research is supported by the DOE grant DE-AC02-83ER40105.

\end{document}